\documentclass[journal]{vgtc}                

\onlineid{1828}

\vgtccategory{Research}

\vgtcpapertype{system}

\title{High-Contrast Projection Mapping under Light Field Illumination\\ with LED Display and Aperiodic Lens Array}

    \author{
      \authororcid{Kotaro Fujimura}{0009-0008-9991-9004},
      \authororcid{Hiroki Kusuyama}{0009-0007-9406-2120},
      \authororcid{Masaki Takeuchi}{0000-0002-0773-7705}, and
      \authororcid{Daisuke Iwai}{0000-0002-3493-5635}
    }
    
    \authorfooter{
      \item
        Kotaro Fujimura, Hiroki Kusuyama, Masaki Takeuchi, and Daisuke Iwai are with the Graduate School of Engineering Science, The University of Osaka.
        E-mail: see \url{https://www.xr.sys.es.osaka-u.ac.jp/}.
    }

\abstract{
Projection Mapping (PM) is a technology that projects images onto the surfaces of physical objects, allowing multiple users to share an augmented reality experience without special devices. However, its practical use has been constrained by the need for dark environments to ensure high-quality projection. To overcome this ``dark-room constraint,'' we propose a novel target-excluding lighting method that selectively illuminates the surrounding environment while avoiding the PM target. Our system achieves light-field illumination by combining an LED display panel with an optimized aperiodic lens array. The key contributions include a compact form factor that provides a large effective light source area, reproducing natural soft shadows comparable to typical lighting, while maintaining the spatial controllability needed to precisely avoid the target. We also introduce a computational technique for optimizing aperiodic lens placement to suppress undesired dark spots caused by crosstalk, and efficient methods for computing LED luminance patterns that enable dynamic PM. Experiments with a prototype system demonstrate that our approach achieves high-contrast PM even in bright environments.
} 


\teaser{
    \centering
    \includegraphics[width=\linewidth]{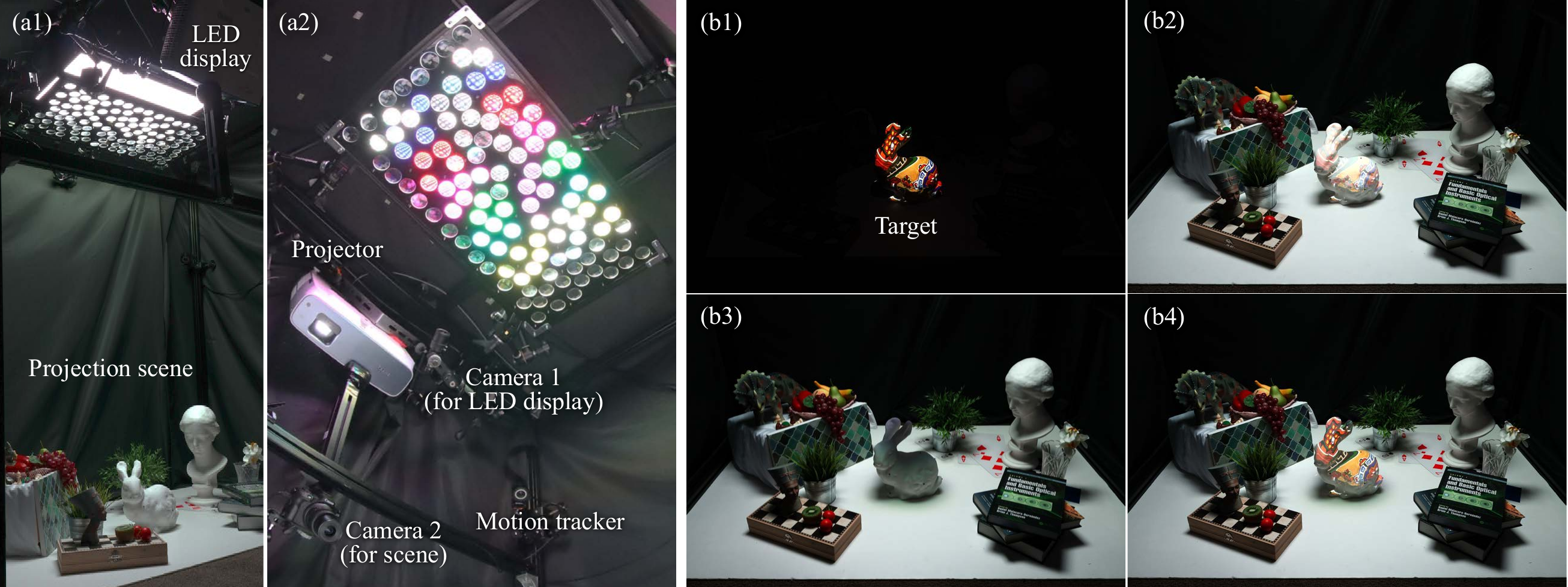}
    \caption{High-contrast projection mapping (PM) in a well-lit environment using the proposed light field illumination. (a1)(a2) Prototype system consisting of an LED display and an aperiodic lens array optimized to suppress unintended dark spots caused by LED-image crosstalk. (b1) PM result in a dark-room environment; (b2) PM result under typical lighting; (b3) scene under target-excluding lighting without PM; (b4) PM result under target-excluding lighting.}
    \label{fig:teaser}
}

\graphicspath{{figures/}{pictures/}{images/}{./}} 

\usepackage{times}                     

\usepackage{tabu}                      
\usepackage{booktabs}                  
\usepackage{lipsum}                    
\usepackage{mwe}                       

\usepackage{mathptmx}                  

\usepackage{subcaption}         
\usepackage{multirow}           
\usepackage{makecell}           
\usepackage[table]{xcolor}      
\usepackage{here}               
\usepackage{bm}
\usepackage{array}
\usepackage{wrapfig}

\usepackage{amsmath} 
\usepackage{amssymb} 

\definecolor{good}{HTML}{d4edda}      
\definecolor{moderate}{HTML}{fff3cd} 
\definecolor{low}{HTML}{f8d7da}      

\newcommand{\argmax}{\mathop{\rm arg~max}\limits}

\begin{document}


\firstsection{Introduction}
\label{sec:INTRODUCTION}

\maketitle

Projection mapping (PM) is a spatial augmented reality (SAR) technique that projects imagery onto the surface of physical objects using a projector, thereby altering their perceived visual properties such as material appearance and color \cite{bimber2005spatial}.
Unlike head-mounted displays, PM does not require users to wear any devices, allowing multiple people to share the same augmented environment with their naked eyes and engage in smooth collaborative interactions.
Its applications extend beyond entertainment \cite{jones2014roomalive} and art \cite{mine2012projection}, covering diverse domains such as intraoperative medical navigation \cite{nishino2018real}, digital makeup \cite{bermano2017makeup,siegl2017faceforge}, urban planning \cite{underkoffler1999urp}, and remote conferencing \cite{raskar1998office,pejsa2016room2room,iwai2017geometrically,fender2017meetalive}.
However, the spread and practical use of PM have long been hindered by the fundamental challenge of interference from environmental illumination.
Projectors can only add light, and thus, under strong ambient lighting, dark components in the projected image appear washed out, and consequently, contrast drops significantly.
To ensure the desired visual quality of PM, it has become almost standard practice to operate under the unrealistic condition of darkening the entire environment.

This ``dark-room constraint'' not only creates operational inconvenience but also introduces two serious problems that undermine the quality of the PM experience itself.
First, when surrounding objects and other viewers sharing the space are obscured in darkness, users find it difficult to perceive their environment.
This directly negates one of PM's fundamental advantages---the ability for multiple people to share an augmented experience with their naked eyes.
Second, and more fundamentally, depending on the content being displayed, the lack of appropriate illumination causes a breakdown of optical consistency between the projected content and the lighting environment \cite{fournier1993common,drettakis1997interactive,debevec2008rendering,sato1999acquiring}.
For instance, to display a specular material such as metal with PM in a physically correct manner, the surrounding environment must be reflected on its surface.
Traditional PM system captures the surrounding environment under bright-room conditions and then maps it onto the object in a dark room.
Yet, since there is no environment to be reflected in the dark room, the resulting imagery appears unnatural and evokes a strong sense of discomfort in viewers.

To address these issues, recent work has proposed replacing conventional illumination with projectors that selectively avoid illuminating the projection target while still lighting the surrounding environment \cite{amano2022reproduction,takeuchi2024projection,yasui2024projection}.
This approach, known as \textit{target-excluding lighting}, is promising because it enables the coexistence of bright environments and high-contrast PM.
Nevertheless, existing projector-based lighting techniques fail to reproduce the natural lighting produced by typical illumination and consequently alter the perceived impression of the scene.
In particular, since a projector approximates a point light source, occluders cast sharp-edged shadows (hard shadows) that differ from the soft shadows produced by area light sources, which are used in typical illumination.
This problem can be mitigated by employing multiple projectors~\cite{amano2022reproduction} or placing a lens array in front of the projector~\cite{yasui2024projection} to synthesize a larger effective light source area, but such solutions inevitably increase system scale and complexity.
As a result, the capability to support dynamic PM, in which projected imagery must track moving targets in real time, remains limited.
Furthermore, typical ambient illumination is intrinsically dominated by low spatial frequencies \cite{belhumeur1998set,ramamoorthi2001relationship,dror2004statistical}.
Consequently, the high resolution of projectors is excessive for lighting purposes and leads directly to impractical drawbacks such as enormous per-pixel computational cost, high power consumption, and substantial heat generation.
Thus, projectors are not an optimal choice of light source for target-excluding lighting.

In this work, we propose a novel lighting technique that overcomes the shortcomings of projector-based approaches.
Our technique combines a low-resolution LED display as the light source with an aperiodic lens array, forming a novel type of integral photography (IP)-based light field illumination \cite{lippmann1908photographie}.
By controlling the LEDs to minimize illumination of the projection target, our system achieves target-excluding lighting that selectively illuminates only the surrounding environment.
The distinctive strength of our approach lies in its ability to achieve both a large effective light source---essential for producing natural illumination with soft shadows---and spatial selectivity to avoid the projection target, all within a compact form factor and with a wide illumination coverage compared to projector-based systems.
In this configuration, the proposed lighting system selectively excludes the projection target while illuminating the surrounding environment, and a standard projector is dedicated solely to PM on the target object, thereby enabling high-contrast PM even under bright ambient conditions.

To realize this framework, we address several technical challenges.
The first contribution concerns the crosstalk of LED images.
In our system, light emitted from each LED passes through multiple lenses in the array, producing not only the primary image but also several secondary images in the scene.
When the lenses are arranged periodically, the spatial luminance pattern of LEDs that are turned off to avoid illuminating the projection target is also repeated, resulting in periodic dark spots across the scene.
We resolve this problem by optimizing the lens arrangement in an aperiodic manner, which disperses these undesirable dark spots and achieves uniform illumination in non-projection regions.
The second contribution addresses the problem of determining the spatial luminance pattern of the LEDs so that only the projection target is excluded from illumination.
Specifically, we develop three techniques:
(1) a method that considers all global illumination effects and produces an optimal LED pattern but requires a long calibration time,
(2) a method that completes calibration in a single shot but cannot account for global illumination effects other than specular reflection and refraction, and
(3) a simplified method that ignores global effects and computes LED patterns in real time.
We evaluate the target-exclusion performance of these approaches using a prototype system and establish usage guidelines for each.
Finally, we demonstrate with a prototype that our system enables high-contrast PM in bright environments, even for dynamic PM scenarios where projection targets move, thereby confirming its effectiveness.

The primary contributions of this work are as follows:
\begin{itemize}
    \item We realize target-excluding lighting that reproduces the natural characteristics of typical illumination while enabling high-contrast PM in bright environments, using a novel architecture that combines an LED display with an aperiodic lens array.
    \item We suppress undesirable dark spots caused by LED image crosstalk by optimizing the aperiodic arrangement of the lens array.
    \item We propose three methods for computing LED luminance patterns that achieve target-excluding lighting.
    \item We evaluate the proposed system through a prototype and demonstrate high-contrast PM in bright environments.
\end{itemize}

\section{Related Work}
\label{sec:RELATED_WORK}

\begin{figure}[t]
    \centering
    \includegraphics[width=\linewidth]{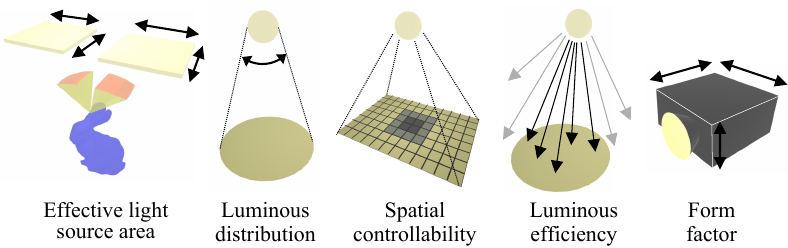}
    \caption{Five evaluation dimensions of spatially selective illumination methods.}
    \label{fig:architecture_comparisonFig}
\end{figure}

\renewcommand{\baselinestretch}{0.7}
\begin{table*}[t]
    \centering
    \caption{Comparison of spatially selective illumination methods and typical environmental lighting.}
    \label{tab:architecture_comparison_modified}
    \resizebox{\textwidth}{!}{

    \begin{tabular}{l|c|c|c|c|c}
        \toprule
        \tiny{Methods} &
        \makecell[c]{\tiny{Effective light}\\ \tiny{source area}} &
        \makecell[c]{\tiny{Luminous} \\ \tiny{distribution}} &
        \makecell[c]{\tiny{Spatial} \\ \tiny{controllability}} &
        \makecell[c]{\tiny{Luminous} \\ \tiny{efficiency}} &
        \makecell[c]{\tiny{Form} \\ \tiny{factor}} \\
        \hline
        
        \makecell[l]{\tiny{Multiple projectors}~\cite{jones2014roomalive, fender2017meetalive, takeuchi2023projection,takeuchi2024projection}} &
        \cellcolor{low}{\tiny{small}} &
        \cellcolor{good}{\tiny{wide}} &
        \cellcolor{good}{\tiny{high}} &
        \cellcolor{moderate}{\tiny{moderate}} &
        \cellcolor{low}{\tiny{bulky}} \\
        
        \makecell[l]{\tiny{Single projector + lens array}~\cite{yasui2024projection, cossairt2008light,yasui2021dynamic,yasui2021wide}} & \cellcolor{moderate}{\tiny{moderate}} &
        \cellcolor{low}{\tiny{narrow}} &
        \cellcolor{good}{\tiny{high}} &
        \cellcolor{low}{\tiny{low}} &
        \cellcolor{moderate}{\tiny{moderate}} \\
        
        \makecell[l]{\tiny{Multiple projectors + lens array}~\cite{zhou2015light,matusik20043d}} &
        \cellcolor{good}{\tiny{large}} &
        \cellcolor{good}{\tiny{wide}} &
        \cellcolor{good}{\tiny{high}} &
        \cellcolor{low}{\tiny{low}} &
        \cellcolor{low}{\tiny{bulky}} \\
        
        \makecell[l]{\tiny{Flat panel display + lens array (w/ mask)}~\cite{takeuchi2016anylight1,takeuchi2016anylight2,takeuchi2018integral, takeuchi2021theory}} &
        \cellcolor{good}{\tiny{large}} &
        \cellcolor{moderate}{\tiny{moderate}} &
        \cellcolor{moderate}{\tiny{moderate}} &
        \cellcolor{low}{\tiny{low}} &
        \cellcolor{good}{\tiny{compact}} \\
        
        \makecell[l]{\tiny{Flat panel display + large-aperture lens}~\cite{kusuyama2024multi}} &
        \cellcolor{good}{\tiny{large}} &
        \cellcolor{low}{\tiny{narrow}} &
        \cellcolor{low}{\tiny{low}} &
        \cellcolor{good}{\tiny{high}} &
        \cellcolor{good}{\tiny{compact}} \\
        
        \makecell[l]{\tiny{Flat panel display + aperiodic lens array (ours)}} &
        \cellcolor{good}{\tiny{large}} &
        \cellcolor{good}{\tiny{wide}} &
        \cellcolor{moderate}{\tiny{moderate}} &
        \cellcolor{moderate}{\tiny{moderate}} &
        \cellcolor{good}{\tiny{compact}} \\
        
        \makecell[l]{\tiny{Typical environmental lighting}} &
        \cellcolor{good}{\tiny{large}} &
        \cellcolor{good}{\tiny{wide}} &
        \cellcolor{low}{\tiny{low}} &
        \cellcolor{good}{\tiny{high}} &
        \cellcolor{good}{\tiny{compact}} \\
        \bottomrule
    \end{tabular}
    }
\end{table*}
\renewcommand{\baselinestretch}{1}

We review prior work that, although not explicitly designed for target-excluding lighting, provides spatially selective illumination.
We compare these approaches along five dimensions (\autoref{fig:architecture_comparisonFig}): \textbf{effective light source area} and \textbf{luminous distribution}, which relate to similarity with typical environmental lighting; \textbf{spatial controllability}, which determines target-exclusion performance; and \textbf{luminous efficiency} and \textbf{form factor}, which relate to practicality.
Effective light source area is defined as the cumulative apparent area of light sources visible from a given point in the scene.
A larger area produces softer shadows and more natural lighting.
Light distribution refers to the angular coverage of illumination.
Typical lighting device, often with diffusers, achieves both a wide effective source area and broad distribution.
Spatial controllability measures how finely illumination can be modulated across space; finer control yields higher values.
Typical environmental lighting has extremely low controllability.
Light efficiency indicates the fraction of emitted light that reaches the environment, while form factor describes the physical footprint of the lighting device.
Typical lighting device is designed to maximize efficiency and minimize bulk to avoid interfering with everyday use.
\autoref{tab:architecture_comparison_modified} summarizes the comparison.

\paragraph{Multiple projectors: }

In the works of RoomAlive \cite{jones2014roomalive}, MeetAlive \cite{fender2017meetalive}, and Takeuchi et al. \cite{takeuchi2023projection, takeuchi2024projection}, the typical environmental lighting is turned off and multiple projectors are distributed throughout the room, transforming the entire space into a single large display, with the projectors simultaneously serving as ambient lighting.
This approach provides extremely high spatial controllability, since the pixels of the projectors can be directly controlled.
In addition, it achieves a wide luminous distribution.
Because light is attenuated inside the projector by the spatial light modulator (SLM) and by the lens aperture, its luminous efficiency is lower than that of ordinary lighting.
Additionally, covering an entire room without blind spots requires many projectors, resulting in poor compactness.
Moreover, since projectors approximate point light sources, their effective source area is small and cast shadows remain hard, unlike the soft shadows of conventional lighting.
Increasing projector count mitigates this but further enlarges and complicates the system.

\paragraph{Projectors and lens array: }

Other studies~\cite{yasui2024projection,zhou2015light,cossairt2008light,yasui2021dynamic,yasui2021wide} apply light field display technologies~\cite{yang2007toward,hirsch2014compressive} that combine a projector with a lens array for lighting.
Here, each lens functions as a sub-projector, increasing the effective source area compared to multiple projectors alone and producing softer, more natural shadows.
Adding more projectors further increases sub-projector density~\cite{zhou2015light,matusik20043d}, but at the cost of system bulk. Covering the full lens array also requires a large projector--lens distance, narrowing the distribution and limiting the luminous distribution~\cite{yasui2021dynamic}.
To address compactness, Yasui et al.~\cite{yasui2021wide} introduced a curved mirror, but this raised new challenges in assembly precision and calibration.
More recently, a kaleidoscope array was proposed~\cite{yasui2024projection} to densify rays and improve shadow softness, though it still suffered from restricted field of view, stray light, block artifacts, and complex calibration.
Overall, optical losses from the projector and lens apertures keep efficiency low.

\paragraph{Flat panel display-based techniques: }

The approach adopted in this research places a lens array in front of a flat-panel display, effectively creating an array of many low-resolution projectors.
This concept has been explored in prior work~\cite{takeuchi2016anylight1,takeuchi2016anylight2,takeuchi2018integral,takeuchi2021theory,zhang2017adjustable}.
Since flat panels act as area light sources, such systems can be thinner than projector-based designs.
However, they suffer from intrinsic crosstalk: light emitted from a single pixel passes through multiple lenses.
This yields a wide source area and broad distribution, but undermines precise spatial control.
Takeuchi et al.~\cite{takeuchi2016anylight1,takeuchi2016anylight2,takeuchi2018integral} mitigated this by inserting an LCD mask to block undesirable rays, which restored spatial controllability but reduced luminous distribution and incurred severe light loss.
Other groups used Fresnel lenses to construct large-aperture projectors, achieving soft shadows through wide effective light source areas~\cite{takeuchi2024projection,kusuyama2024multi}.
However, these systems have shallow depth of field, resulting in poor spatial controllability for target-excluding lighting of 3D objects.

\paragraph{Our Contribution: }

In this study, we implement target-excluding lighting using an LED array with a lens array.
As discussed above, the key challenge of this design is loss of spatial controllability due to crosstalk.
With periodic lens arrangements, crosstalk images overlap periodically, causing undesirable dark spots when LEDs are turned off to avoid the projection target.
We propose an aperiodic optimization of lens placement that disperses crosstalk images spatially, enabling uniform illumination of non-target regions.
This achieves high spatial controllability in target-excluding lighting without introducing light loss beyond lens apertures, while preserving compactness and broad luminous distribution.
As summarized in \autoref{tab:architecture_comparison_modified}, our method provides a promising balance of properties compared to existing designs.

\section{Method}
\label{sec:METHOD}

To achieve high-contrast PM in a well-lit environment, we propose a new light-field illumination technique that combines an LED display panel with an aperiodic lens array.
The key idea is to selectively avoid direct illumination on the projection target area while lighting only the surrounding environment.
By suppressing illumination that would otherwise wash out the projected imagery, the proposed approach maintains high contrast even under ambient lighting.
This section details two technical components essential to our method.
First, we optimize an aperiodic arrangement of lenses to maximize illumination uniformity in the non-target area and thereby suppress undesirable dark spots that arise during target-excluding lighting (Section \ref{sec:LensArrayOptimization}).
Second, we compute spatial luminance patterns for the LED display that avoid illuminating the target area (Section \ref{sec:LedSelection}).

\subsection{Optimization of an Aperiodic Lens Array}
\label{sec:LensArrayOptimization}

\begin{figure}[t]
    \centering
    \includegraphics[width=\linewidth]{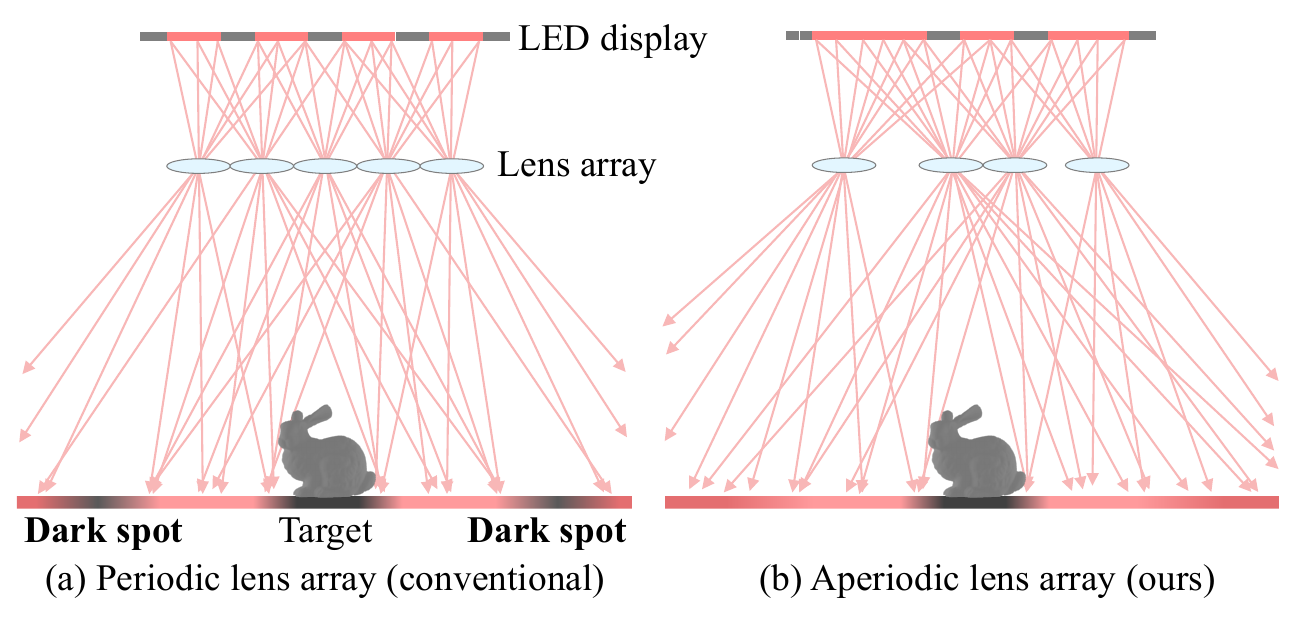}
    \caption{Crosstalk-induced dark spots in target-excluding lighting with a conventional periodic lens array (a), and their suppression with the proposed aperiodic lens array (b).}
    \label{fig:LightFieldIllumi_ConceptFig}
\end{figure}

When pixels that illuminate the target are turned off, not only the rays passing through the nearest lens vanish, but rays passing through neighboring lenses also disappear.
This crosstalk produces unintended dark spots in surrounding areas.
If the lens array is laid out periodically, aliasing induces low-frequency brightness fluctuations with periods longer than the lens pitch, severely degrading illumination quality (\autoref{fig:LightFieldIllumi_ConceptFig}(a)).
We therefore break periodicity and spatially disperse crosstalk images to mitigate aliasing (\autoref{fig:LightFieldIllumi_ConceptFig}(b)).

\begin{figure}[tb]
    \centering
    \includegraphics[width=\linewidth]{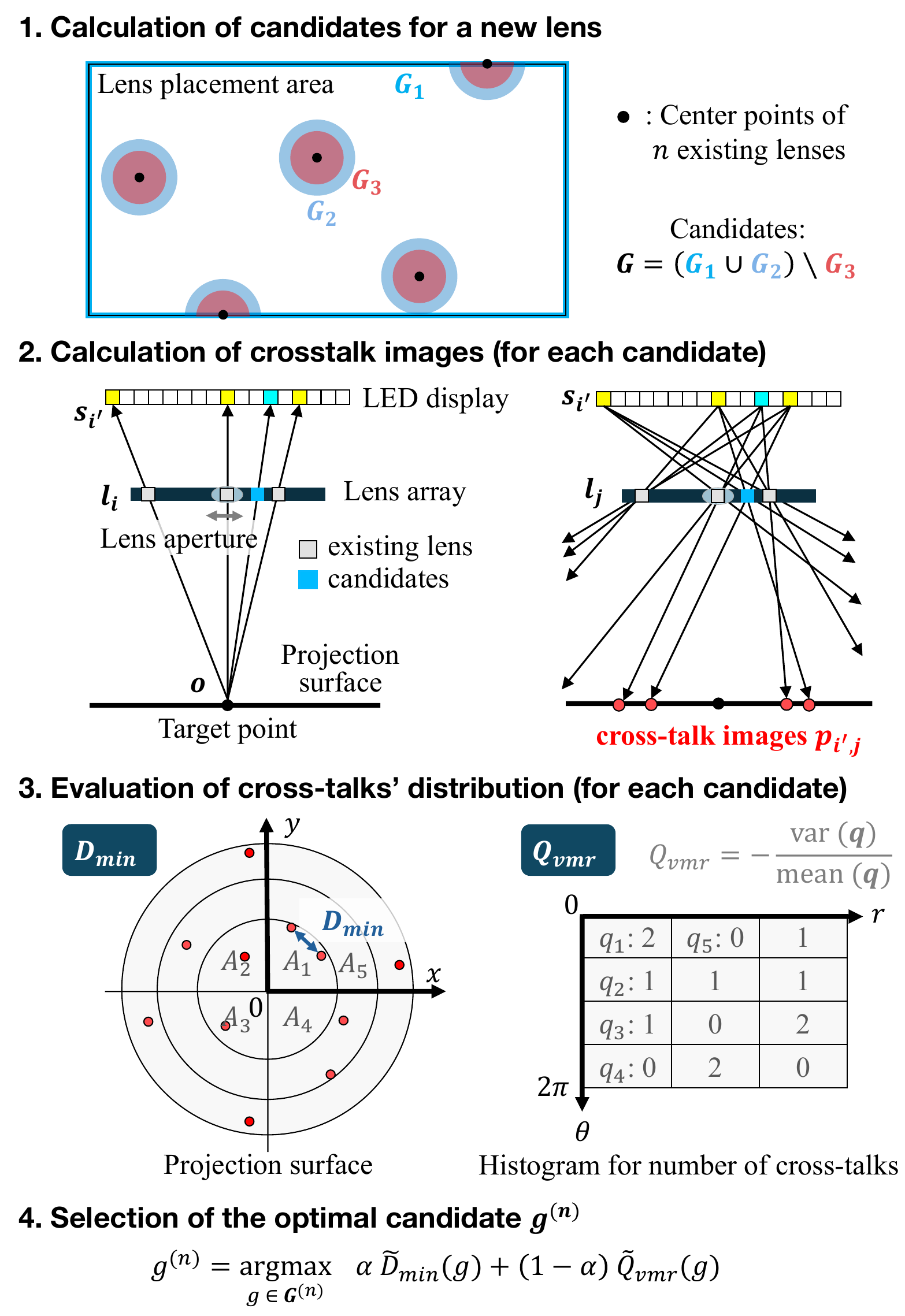}
    \caption{Optimization process of lens placement.}
    \label{fig:LensOpti_ConceptFig}
\end{figure}

We adopt a greedy algorithm to search for an effective layout.
Starting from a configuration with $n\!-\!1$ lenses, we compute multiple candidate lens positions for the $n$-th lens and select the position that yields the most uniform irradiance over the scene.
This iteration continues until no further lenses can be placed physically.
To suppress aliasing and promote uniformity, crosstalk images of LED pixels must be well dispersed.
Accordingly, we choose lens positions that maximize both the nearest-neighbor distance among crosstalk images and the global spread of their locations.
We define a coordinate system whose origin lies at the center of the LED display.
The display coincides with the $xy$-plane, and its normal is along $+z$.
Lenses are placed on a rectangular plane, parallel to the LED panel and comparable in size, which is translated by $z_{lens}$ along the $z$-axis.
We refer to this rectangle as the \emph{placement region}.
\autoref{fig:LensOpti_ConceptFig} visualizes the whole optimization process.

\subsubsection{Computing Candidate Lens Positions}\label{sec:OptimizationProcess}

Let $\bm{G}_{all}$ denote the set of discrete grid points defined inside the placement region.
The optimal lens position is searched from this set of grid points.
To reduce computational cost and enable efficient exploration, we prune the grid points and restrict the candidate positions.
Specifically, given a configuration where $n\!-\!1$ lenses have already been placed, the position of the $n$-th lens is searched from a subset of grid points defined as
\[
\bm{G}^{(n)} = \left(\bm{G}_1^{(n)} \cup \bm{G}_2^{(n)}\right) \setminus \bm{G}_3^{(n)}.
\]
Here, $\bm{G}_1^{(n)}$ is the set of all grid points located on the boundary $C$ of the placement region;
$\bm{G}_2^{(n)}$ is the set of all grid points contained within a region $S$ defined as the area within a distance $r_{max}$ from the centers of the already placed lenses;
and $\bm{G}_3^{(n)}$ is the set of all grid points contained within a region $S'$ where additional placement is physically impossible due to overlap with existing lenses.
The region $S'$ is defined as the area within a distance of $r_{lens} + \delta$ from each lens center, where $r_{lens}$ is the lens radius and $\delta$ is a small margin introduced to prevent adjacent lenses from touching.

\subsubsection{Computing Crosstalk Image Locations}\label{sec:EfficientCrostalkCalculation}

To evaluate lens positions, it is necessary to compute the distribution of crosstalk images.
We assume a pinhole model.
The crosstalk images are evaluated on a plane parallel to the LED display, located at $z=z_{proj}$, which we refer to as the \textit{evaluation plane}.
Let the projection target be a point at $\bm{o}=(0,0,z_{proj})$ on the evaluation plane.
For the $i$-th lens positioned at $(x_i, y_i, z_{lens})$ $(i=1,2,\ldots,n)$, the $xy$-coordinate $\bm{s}_{i'} \in \mathbb{R}^2$ of the LED pixel $i'$ that illuminates the target through this lens is given by
\begin{equation}
    \bm{s}_{i'} = \bm{l}_i + (\bm{o}-\bm{l}_i)\frac{z_{src} - z_{lens}}{z_{proj} - z_{lens}},
    \label{eq:inverse_ray_tracing_marker}
\end{equation}
where $z_{src}\ (=0)$ denotes the $z$-coordinate of the LED pixels, and $\bm{l}_i \in \mathbb{R}^2$ represents the $xy$-coordinate $(x_i, y_i)$ of lens $i$.
Next, consider the light emitted from LED pixel $\bm{s}_{i'}$.
When this light passes through another lens $j~(j=1,2,\ldots,n;\; j \neq i)$, the resulting crosstalk image formed on the evaluation plane at $z=z_{proj}$ has $xy$-coordinate $\bm{p}_{i',j} \in \mathbb{R}^2$ given by
\begin{equation}
    \bm{p}_{i',j} = \bm{l}_j + (\bm{l}_j - \bm{s}_{i'})\frac{z_{proj} - z_{lens}}{z_{lens} - z_{src}}.
    \label{eq:forward_ray_tracing}
\end{equation}
By performing this computation for all LED pixels $\{i'\}$ that illuminate the target point $\bm{o}$, and for all other lenses $\{j\}$ with $j \neq i$, we obtain the complete set of $n(n-1)$ crosstalk image positions, $\bm{P} = \{\bm{p}_{i',j}\}$.

\subsubsection{Optimal Lens Placement for Uniform Crosstalk Distribution}
\label{sec:EvaluationFunction}

For each candidate grid point $g$ in $\bm{G}^{(n)}$, we compute the distribution $\bm{P}$ of crosstalk images on the evaluation plane using the method described in Section \ref{sec:EfficientCrostalkCalculation}, and evaluate the spatial uniformity of this distribution.
The candidate point $g^{(n)}$ that yields the highest evaluation score is chosen as the $n$-th lens position.
As the evaluation metric, we introduce the weighted sum $E(g)$ of two factors: the smallest Euclidean distance $D_{min}(g)$ among all pairs of crosstalk images in $\bm{P}$, and the variance-to-mean ratio score $Q_{vmr}(g)$ of the overall crosstalk distribution.
Specifically, for each candidate grid point, we compute $D_{min}(g)$ and $Q_{vmr}(g)$, normalize them to the range $[0,1]$ as $\tilde{D}_{min}(g)$ and $\tilde{Q}_{vmr}(g)$, and then determine the optimal position as
\begin{eqnarray}
    g^{(n)}&=&\argmax_{g\in\bm{G}^{(n)}}E(g),\\
    E(g) &=& \alpha\tilde{D}_{min}(g) + (1-\alpha)\tilde{Q}_{vmr}(g),
    \label{eq:s_score}
\end{eqnarray}
where $\alpha$ $(0 \leq \alpha \leq 1)$ is a weighting factor.
For brevity, we omit $g$ in the following discussion.

\paragraph{Minimum distance of crosstalk images:}
$D_{min}$ is defined as the smallest Euclidean distance among all pairs of crosstalk images in $\bm{P}$.
A larger $D_{min}$ indicates that the crosstalk images are more evenly distributed without clustering.
Let $\bm{p}_k,\bm{p}_l~(k,l=1,2,\ldots,n(n-1))$ be elements of $\bm{P}$. Then,
\begin{subequations}\label{eq:lmin}
    \begin{equation}
        D_{min} = \min_{k,l} ~ D_{k,l},
    \end{equation}
    \begin{equation}
        D_{k,l} = ~
        \begin{cases}
            ~~\| \bm{p}_k - \bm{p}_l \|_2 &~~ \text{if} ~~ k \neq l , \\
            ~~\infty &~~ \text{otherwise},\\
        \end{cases}
    \end{equation}
\end{subequations}
Computing $D_{k,l}$ requires considering all pairs of crosstalk images, which incurs a complexity of $\mathcal{O}(n^4)$ at each step of the greedy optimization.
To reduce this computational burden, we propose a recursive computation that leverages the greedy process.
Specifically, when adding the $n$-th lens, we only compute distances involving the newly generated crosstalk images, omitting redundant calculations among the existing ones.
Let $\bm{P}^{(n-1)}$ denote the set of crosstalk images produced by $n-1$ lenses, with a known minimum distance $D_{min}^{(n-1)}$.
When the $n$-th lens is added, the new set of crosstalk images is $\bm{P}^{(new)}=\bm{P}^{(n)}\setminus\bm{P}^{(n-1)}$.
The updated minimum distance is then given by
\begin{equation} \label{eq:prop_dmin}
    D_{min}^{(n)} = \min\{D_{min}^{(n-1)}, D_{min}^{(new)}, D_{min}^{(new\text{-}prev)}\},
\end{equation}
where $D^{(new)}$ and $D^{(new\text{-}prev)}$ denote the minimum pairwise distances among $\bm{P}^{(new)}$ and between $\bm{P}^{(new)}$ and $\bm{P}^{(n-1)}$, respectively, defined as
\begin{subequations}\label{eq:Dnew}
    \begin{equation}\label{eq:Dnew1}
        D^{(new)}_{min} = \min_{k,l} ~ D^{(new)}_{k,l}, ~~D^{(new\text{-}prev)}_{min} = \min_{k,l} ~ D^{(new\text{-}prev)}_{k,l},
    \end{equation}
    \begin{equation}\label{eq:Dnew2}
        D^{(new)}_{k,l} = ~
        \begin{cases}
            ~~\| \bm{p}_k - \bm{p}_l \|_2~~ (\bm{p}_k, \bm{p}_l \in \bm{P}^{(new)}) &~~ \text{if} ~~ k \neq l , \\
            ~~\infty &~~ \text{otherwise}, \\
        \end{cases}
    \end{equation}
    \begin{equation}\label{eq:Dnew3}
        D^{(new\text{-}prev)}_{k,l} = ~
        \begin{cases}
            ~~\| \bm{p}_k - \bm{p}_l \|_2~~ (\bm{p}_k \in \bm{P}^{(new)},~\bm{p}_l \in \bm{P}^{(n-1)}) &~~ \text{if} ~~ k \neq l , \\
            ~~\infty &~~ \text{otherwise}. \\
        \end{cases}
    \end{equation}
\end{subequations}
Note that $|\bm{P}^{(new)}| = n(n-1) - (n-1)(n-2) = 2n-2$ and $|\bm{P}^{(n)}| = n(n-1)$.
Accordingly, \autoref{eq:Dnew2} requires $(2n-2)\times(2n-2)$ computations, while \autoref{eq:Dnew3} requires $(2n-2)\times n(n-1)$ computations.
Thus, the computational complexity of this recursive approach is $\mathcal{O}(n^3)$, which is smaller than the $\mathcal{O}(n^4)$ complexity of the naive method that directly computes all $D_{k,l}$ values.

\paragraph{Spatial distribution:}
To assess the spatial distribution of crosstalk images, we employ the Variance-to-Mean Ratio (VMR) score $Q_{vmr}$.
The convex hull $A$ enclosing all crosstalk images on the evaluation plane is divided into $N$ sectors $(A_1, A_2, \ldots, A_N)$ in polar coordinates.
Let $q_i$ denote the number of crosstalk images within sector $A_i$.
The VMR is then defined as the ratio of the variance to the mean of $\{q_i\}$.
A smaller VMR indicates a more uniform distribution across the evaluation region.
To fit the maximization framework, we define the score as $Q_{vmr} = -VMR$:
\begin{subequations}\label{eq:q_score}
    \begin{equation}
        Q_{vmr} = -VMR = - \frac{\text{var}(\bm{q})}{\text{mean}(\bm{q})},
    \end{equation}
    \begin{equation}
        \bm{q} = \{q_i\}~(i=1,2,\ldots,N),~~ q_i = |\{\bm{p} \subset A_i\}|,
    \end{equation}
\end{subequations}

\subsection{LED Intensity Pattern Computation for Target-Excluding Lighting}
\label{sec:LedSelection}

To selectively avoid illuminating the projection target while lighting only the surrounding scene, we design three methods for computing LED pixel intensity patterns, each with distinct characteristics.
Specifically, we propose: (1) a method that provides an optimal LED pattern accounting for all global illumination effects but requires prohibitively long calibration time (Section \ref{sec:LedSelection-LTM}); (2) a method that completes calibration in a single shot but cannot account for global effects other than specular reflection and refraction (Section \ref{sec:LedSelection-RT}); and (3) a method that ignores global illumination and considers only direct illlumination components, enabling real-time computation at the cost of reduced accuracy (Section \ref{sec:LedSelection-Sim}).

\subsubsection{Light Transport Matrix-based Method}
\label{sec:LedSelection-LTM}

When light from LED pixels propagates into the scene, not only the directly illuminated points but also other points become bright due to global illumination effects such as interreflection, subsurface scattering, specular reflection, and refraction.
As a result, the projection target may still appear bright even if it is not directly lit by the LEDs.
In PM research, techniques have been developed to compensate for global illumination effects in order to reproduce the desired appearance.
Let $\bm{a}$ denote the vector of input values $a_i$ for all LED pixels, and let $\bm{b}$ denote the vector of observed pixel values $b_j$ from a camera.
Using the light transport matrix (LTM) $\bm{T}$, the relation $\bm{b}=\bm{T}\bm{a}$ holds~\cite{sen2005dual}.
Thus, to reproduce a desired appearance $\bm{b}'$, the illumination pattern $\bm{a}'$ can be obtained as $\bm{a}'=\bm{T}^{-1}\bm{b}'$~\cite{4392749}.
Since the LTM captures all global illumination effects, this method theoretically yields an LED pattern that minimizes unintended illumination on the target.

Although originally developed for a projector–camera pair, we generalize this approach to our system.
Even with lens arrays or in the presence of complex phenomena such as crosstalk, the relationship between an array of emitters (LEDs) and receivers can still be modeled with an LTM.
Here, $\bm{a}$ corresponds to the LED pixel intensities.
This method can therefore minimize the influence of global illumination and remains effective even when the geometry of the projection target is unknown, provided the target region can be segmented in camera images.
The drawbacks are twofold.
First, measuring the LTM requires an extremely long acquisition time, on the order of tens of hours.
Second, since the inverse $\bm{T}^{-1}$ may produce negative or out-of-range values, the LED pattern $\bm{a}'$ cannot be computed by direct multiplication.
Instead, iterative optimization such as gradient descent is used, resulting in high computational cost.

\subsubsection{Optical Ray Tracing Method}
\label{sec:LedSelection-RT}

While the LTM-based method provides an optimal LED pattern for target-excluding lighting, its measurement and optimization costs are prohibitively high.
As an alternative, we propose a simpler approach that identifies target-illuminating LED pixels through one-shot calibration, and then illuminates only the remaining pixels to light the surrounding environment.
This method is inspired by the ray tracing algorithm in computer graphics, in which rays are traced backward from camera pixels to the light source.
In our technique, we determine the target-illuminating pixels through the following procedure.
First, as an offline step, we obtain the geometric correspondence between the camera and the LED display pixels in advance.
Next, with the LED display turned off, we project white light from the projector onto the projection target.
The light reflected by the object passes through the lens array and forms an image on the LED display surface.
We capture this image with the camera and identify ``bright'' pixels via thresholding.
Based on Helmholtz reciprocity, the LED pixels observed as bright correspond to those that would illuminate the target object if turned on.
Finally, we convert these pixel locations into LED indices using the C2P map and turn them off to generate the target-excluding illumination pattern.
Calibration requires only a single camera capture of the LED panel under white projection, making the process extremely fast.
Moreover, similar to ray tracing in graphics, this method accounts not only for direct illumination components but also for global effects due to mirrors and refractive objects, thereby enhancing target exclusion performance.
As with the LTM-based method, it also does not require knowledge of the target geometry, as long as the target region can be segmented in captured images.
In implementation, a binary threshold is applied to classify the bright pixels.
Since the optimal threshold depends on the hardware and environmental lighting, it is determined empirically during calibration to ensure that reflections from the target are detected while weak scattered light is ignored.
Note that measurement errors may arise from geometric correspondence estimation, typically on the order of a few pixels. However, this margin of error is well within an acceptable range for the purpose of target-excluding lighting, as it does not critically affect visual quality.

\subsubsection{Geometry-based Method}
\label{sec:LedSelection-Sim}

The two methods above require projecting structured patterns whenever the scene changes, which makes them unsuitable for dynamic PM where the target moves.
To address this, we propose a faster method that ignores global illumination and considers only direct illumination components.
Although this reduces accuracy, it enables calibration-free, real-time computation of LED patterns for target-excluding lighting.

In this method, a motion tracking system is used to acquire the 3D coordinates of markers attached to the target surface in real time.
From each marker, virtual rays are traced through the centers of all lenses, and their intersections with the LED panel determine the corresponding LED pixels.
For each lens, the convex hull of these intersected pixels is computed.
Target-illuminating pixels are defined as those within the convex hulls, and by turning on only the pixels outside them, target-excluding lighting is achieved.
Because the number of markers is relatively small (on the order of a dozen), the computational cost is low, allowing real-time operation at frame rates above 60 Hz.
However, there is a trade-off: increasing the number of markers slows down processing but yields convex hulls that better approximate the target geometry, thereby improving exclusion performance.
Conversely, fewer markers allow faster processing but reduce accuracy in target exclusion.

\section{Experiment}

\subsection{Prototype}\label{sec:IMPLEMENTATION_ENVIRONMENT}

An overview of the experimental system is shown in \autoref{fig:teaser}(a).
The system consists of an LED display serving as the light source, a lens array, a projector for PM onto the target object, cameras, and a motion-tracking system for estimating the target's pose.
All devices are controlled from a single PC.

As the light source, we used an LED display panel (Barco XT2.5-HB).
Its key specifications are a 2.54\,mm pixel pitch, a resolution of 240\,$\times$\,135 pixels, and a maximum luminance of 1500\,nits.
The lens array is composed of lenses with a 38\,mm aperture, 2\,mm edge thickness, and 100\,mm focal length.
We fabricated the array by drilling holes in a black acrylic plate (730\,mm\,$\times$\,390\,mm) and inserting the lenses.
For PM onto the Stanford bunny target (187\,mm\,$\times$\,229\,mm\,$\times$\,229\,mm), we used a DLP projector (BenQ TK850).
Two cameras (Canon EOS R50 and POINTGREY Flea3 FL3-U3-32S2C) were used for system calibration and evaluation: one for LTM measurement (Section \ref{sec:LedSelection-LTM}) and capturing the overall projection surface, and the other for imaging the LED display in the optical ray tracing method (Section \ref{sec:LedSelection}).
For tracking of the target movement, four OptiTrack Flex13 units were placed at the corners of the setup.
We attached 18 hemispherical retro-reflective markers (diameter: approx. 3 mm) distributed evenly across the surface of the projection target.
As adopted in previous work \cite{nomoto2022dynamic}, this setup minimizes visual interference for observers since the markers reflect light primarily toward the tracking cameras and are sufficiently small relative to the projection target so as not to detract from its visual appearance.
The controlling PC was equipped with an Intel Core i7-9700 (3.00\,GHz) CPU, an NVIDIA GeForce RTX 3060 GPU, and 16.0\,GB RAM.
The geometric calculations for the geometry-based method were implemented on the GPU using CUDA to ensure real-time performance.
The experimental environment was assembled on a floor area of 1000\,mm\,$\times$\,1400\,mm.
The LED panel was mounted 1650\,mm above the floor, facing downward; the lens array was installed 110\,mm below the LED panel, i.e., at a height of 1540\,mm.

\subsection{Aperiodic Lens-Array Optimization and Suppression of Crosstalk-Induced Dark Spots}
\label{sec:IMPLEMENTATION_LENSARRAY}

\begin{figure}[t]
    \centering
    \includegraphics[width=\linewidth]{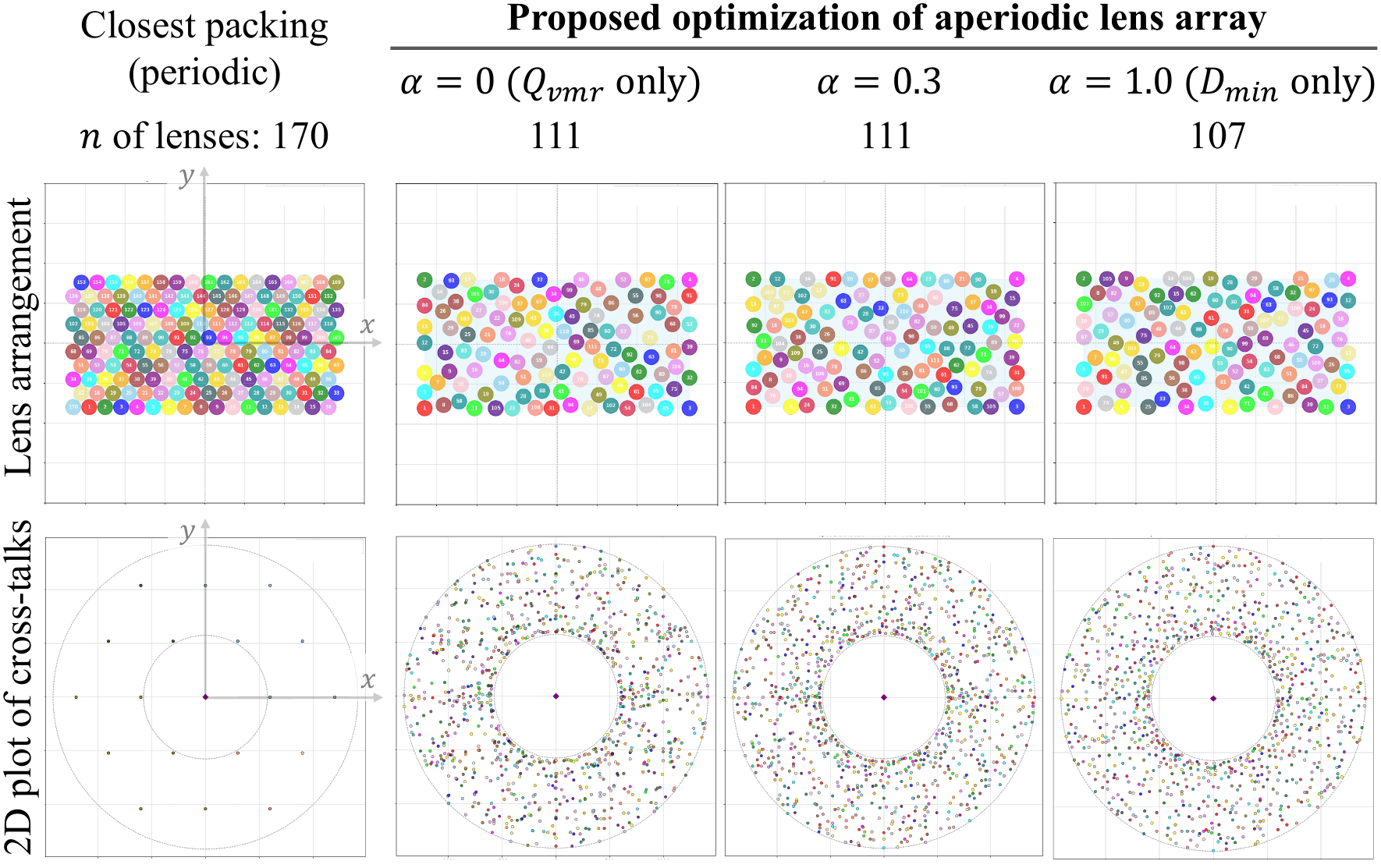}
    \caption{Optimization results of lens placement and the corresponding 2D distributions of crosstalk images.}
    \label{fig:simulation}
\end{figure}

We designed the aperiodic lens-array using the proposed optimization method.
We set the grid pitch to 0.5 mm and the lens placement region to $x \in [-330 \text{ (mm)}, 330 \text{ (mm)}], y \in [-160 \text{ (mm)}, 160 \text{ (mm)}]$.
The size of the discrete grid point set $G_{all}$ is 846,761.
The first four lenses were placed at the corners of the rectangular placement region.
Empirically, starting from the periphery allows a larger number of lenses to be placed than starting from the interior.
We then conducted independent optimizations with different weighting parameters $\alpha$ in \autoref{eq:s_score}, using step increments of 0.1 ($0 \le \alpha \le 1$).
Since the closest packing arrangement is periodic and yields the highest luminous efficiency, we adopt it as the baseline.
\autoref{fig:simulation} shows the baseline lens arrangement and the optimized ones together with the 2D plots of the crosstalk images for $\alpha=0.0, 0.3,$ and $1.0$.
Although the baseline packed 170 lenses, its crosstalk images cluster periodically on the evaluation plane, suggesting that periodic dark spots would emerge during target-excluding lighting.
In contrast, our method yielded approximately 
111 
lenses for all weight settings, yet the crosstalk images were evenly dispersed.
We also measured computation time.

With the na\"{i}ve computation of $D_{min}$ (\autoref{eq:lmin}), our implementation required about 6.5 hours to complete placement.
Using the proposed recursive update (\autoref{eq:prop_dmin}, \autoref{eq:Dnew}), the total time was reduced to about 0.75 hours.

\begin{figure}[t]
    \centering
    \includegraphics[width=\linewidth]{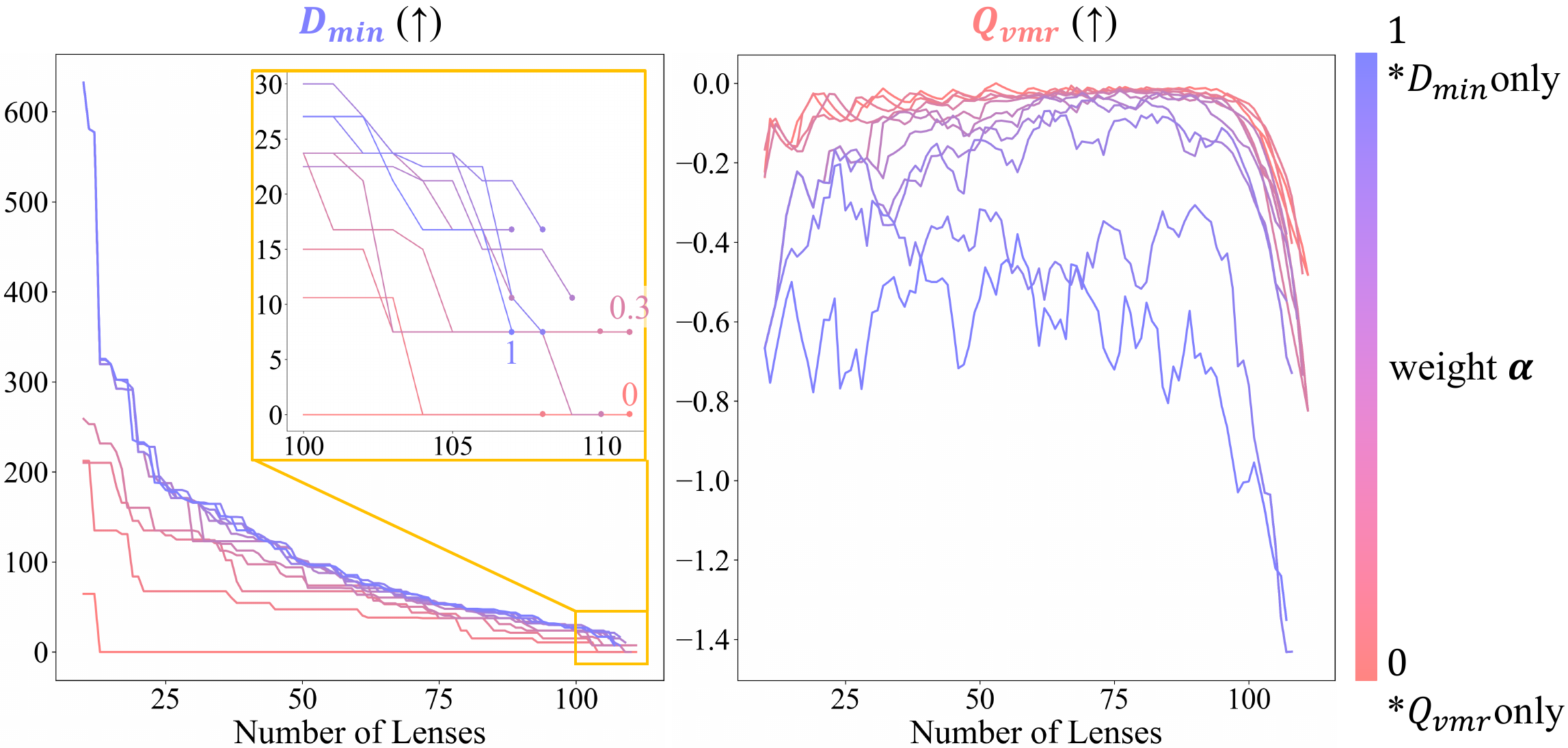}
    \caption{$D_{min}$ and $Q_{vmr}$ as the number of lenses increases, under different weight parameters $\alpha$. In both metrics, a higher value indicates a better evaluation. The closer $\alpha$ is to 1 (blue), the greater the emphasis on $D_{min}$; the closer it is to 0 (red), the greater the emphasis on $Q_{vmr}$.}
    \label{fig:dmin_qvmr}
\end{figure}

\autoref{fig:dmin_qvmr} plots $D_{min}$ and $Q_{vmr}$ as functions of the number of placed lenses under different $\alpha$.
As $\alpha$ increases, $D_{min}$ improves; conversely, decreasing $\alpha$ (increasing the weight on $Q_{vmr}$) improves $Q_{vmr}$.
More specifically, when $\alpha$ is small, increasing the number of lenses leads to cases where $D_{min}=0$, meaning that crosstalk images overlap.
This corresponds to the observation in Figure~5 for $\alpha=0$, where the number of crosstalk image points appears smaller than in other cases despite the lens count being comparable to the other conditions, indicating that multiple images are overlapping.
In contrast, when $\alpha$ is large, the optimization prioritizes maximizing $D_{min}$, which consequently reduces the final number of lenses that can be placed.
Because the number of lenses directly affects the effective light source area and overall brightness, we selected $\alpha=0.3$ for fabricating the physical prototype used in subsequent experiments. This value not only yielded the largest lens count (111 lenses) but also ensured $D_{min} \neq 0$, meaning that no crosstalk images overlapped.

\begin{figure}[t]
    \centering
    \includegraphics[width=\linewidth]{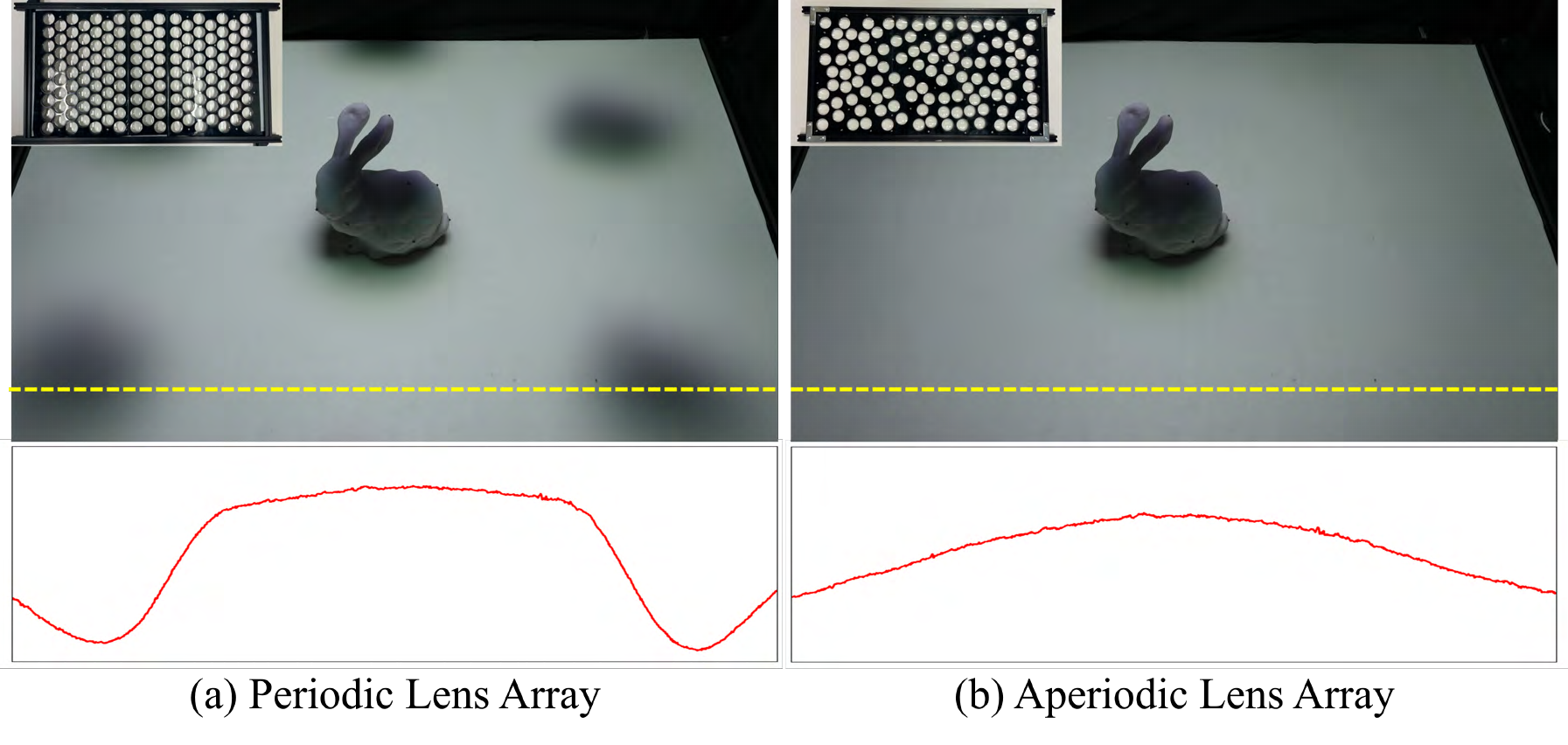}
    \caption{Comparison of unintended dark spots in illuminated scene using (a) the periodic lens array and (b) the optimized aperiodic lens array. The figure below shows the luminance intensity profile along the dotted line in the figure above.}
    \label{fig:artifact_suppression_result}
\end{figure}

Next, we fabricated an aperiodic array with $\alpha=0.3$ and a baseline closest packing array to quantitatively evaluate the effect of the proposed optimization on suppressing undesirable dark spots during target-excluding lighting.
Because of manufacturing constraints, the baseline periodic array was fabricated with 146 lenses (rather than 170).
In the experiment, the target was placed at the center of the floor, and target-excluding lighting was performed.
The LED's spatial luminance pattern was computed with the optical ray tracing method (Section \ref{sec:LedSelection-RT}).
We then captured the scene under target-excluding lighting for each array.
\autoref{fig:artifact_suppression_result} show the results for the periodic and aperiodic arrays, respectively.
The periodic array exhibits periodic dark spots around the non-illuminated region, whereas the aperiodic array does not show such periodic artifacts.
These results confirm that the proposed aperiodic lens arrangement effectively disperses crosstalk-induced dark spots and achieves uniform illumination in non-target regions.


\subsection{Projection Mapping under Target-Excluding Lighting}\label{sec:EXPERIMENT_PM}

\begin{figure}[t]
    \centering
    \includegraphics[width=\linewidth]{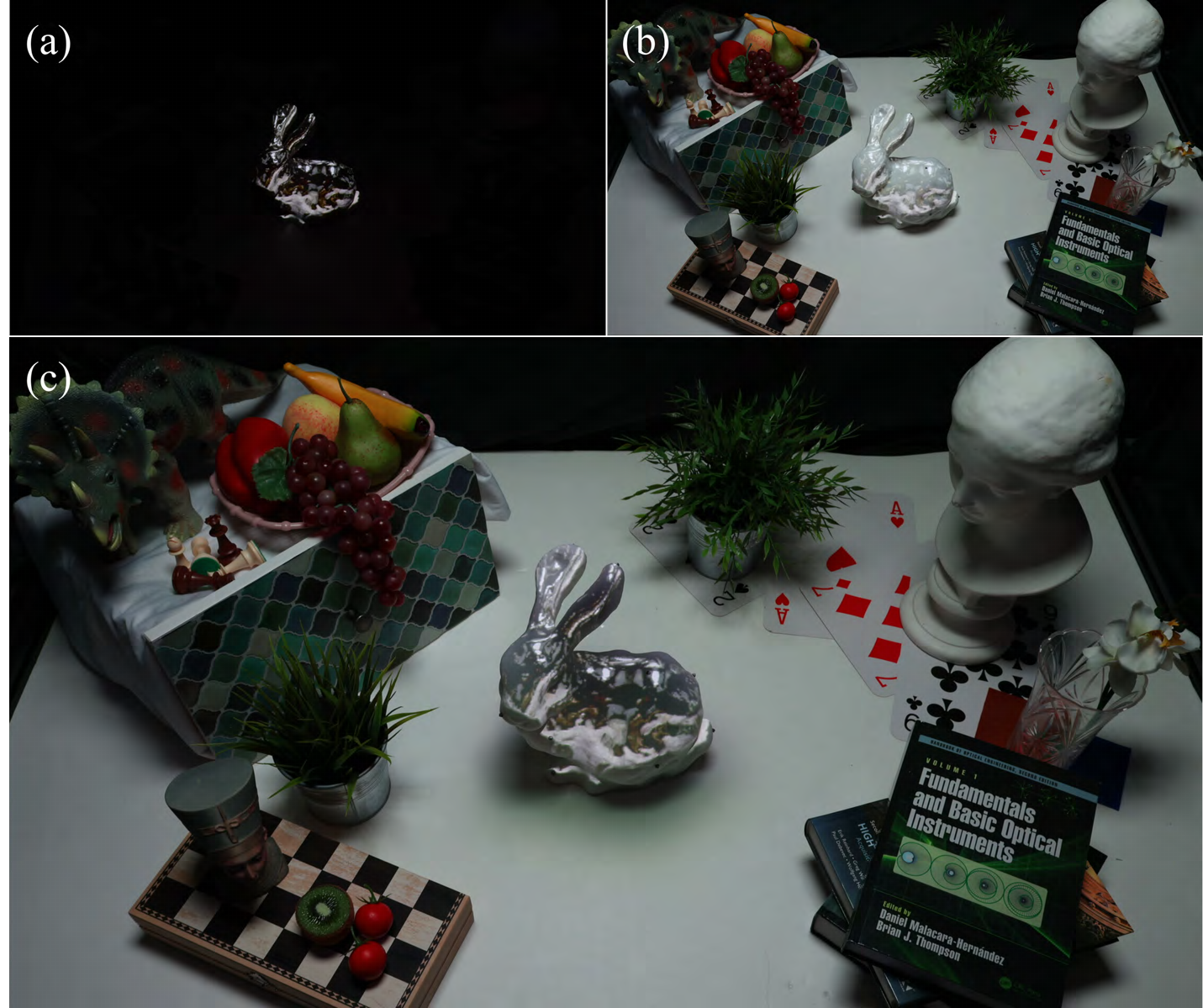}
    \caption{PM of specular material representation under (a) dark, (b) bright environments, and (c) target-excluding lighting.}
    \label{fig:specular_PM}
\end{figure}

Using the system with the aperiodic lens array at $\alpha=0.3$ shown in \autoref{fig:teaser}(a), we implemented PM under the proposed target-excluding lighting.
We placed the Stanford bunny at the center of the floor and arranged multiple objects around it.
\autoref{fig:teaser}(b) and \autoref{fig:specular_PM} show PM results under three conditions: a dark room, a typical bright room (all LED pixels on), and target-excluding lighting with the LED pattern computed by the optical ray tracing method.
In the dark room, the PM appears high-contrast and crisp, but the surrounding objects are not visible.
In the bright room, the surroundings are visible but PM contrast is degraded.
With our method, the surroundings remain visible while PM contrast remains high.
In particular, when rendering specular content (as in \autoref{fig:specular_PM}), our method is essential.
Although the dark-room condition yields high-contrast and photorealistic specular highlights, the absence of visible surroundings breaks optical consistency and reduces physical realism.
Our method preserves high-contrast specular reflections with visible surroundings, maintaining optical consistency and realism.

\begin{figure*}[t]
    \centering
    \includegraphics[width=\linewidth]{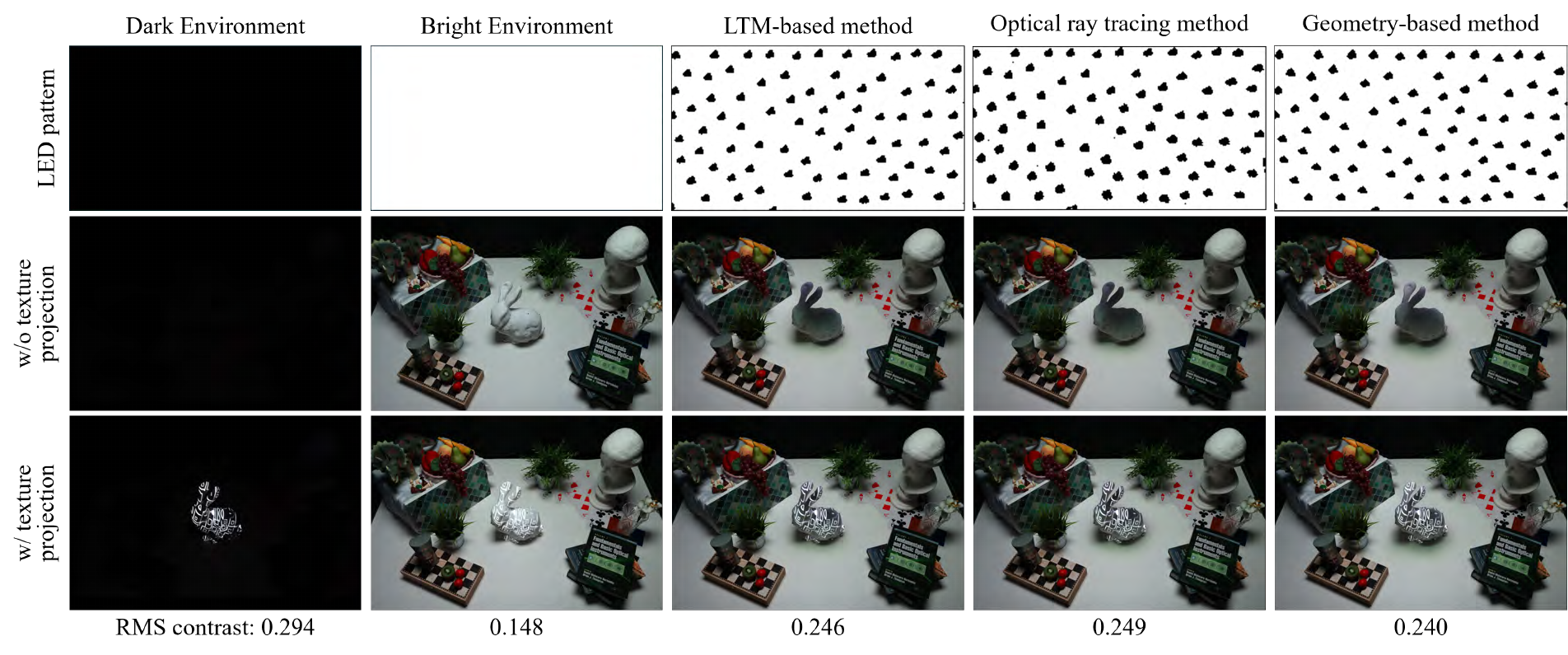}
    \caption{Quantitative comparison of PM contrast under five lighting conditions. Each condition shows the LED luminance pattern (top), pre-projection scene (middle), and projected result with RMS contrast values (bottom).}
    \label{fig:pm_results_Without_Mirror}
\end{figure*}

We next measured PM contrast quantitatively.
A binary black and white texture was projected, and contrast was computed under five conditions: dark room, bright room with all LEDs on, and three target-excluding conditions corresponding to the three LED-pattern computation methods.
For each condition, \autoref{fig:pm_results_Without_Mirror} shows the LED luminance pattern, the pre-projection scene, and the projected result together with the RMS contrast over the PM target region.
The results reveal three findings.
First, with our method, contrast remained at least 80 \% of the dark-room value in all cases.
Second, contrast was at least 1.6 times higher than that of the bright-room condition.
Third, differences among the three LED pattern computation methods were small.

\begin{figure}[t]
    \centering
    \includegraphics[width=\linewidth]{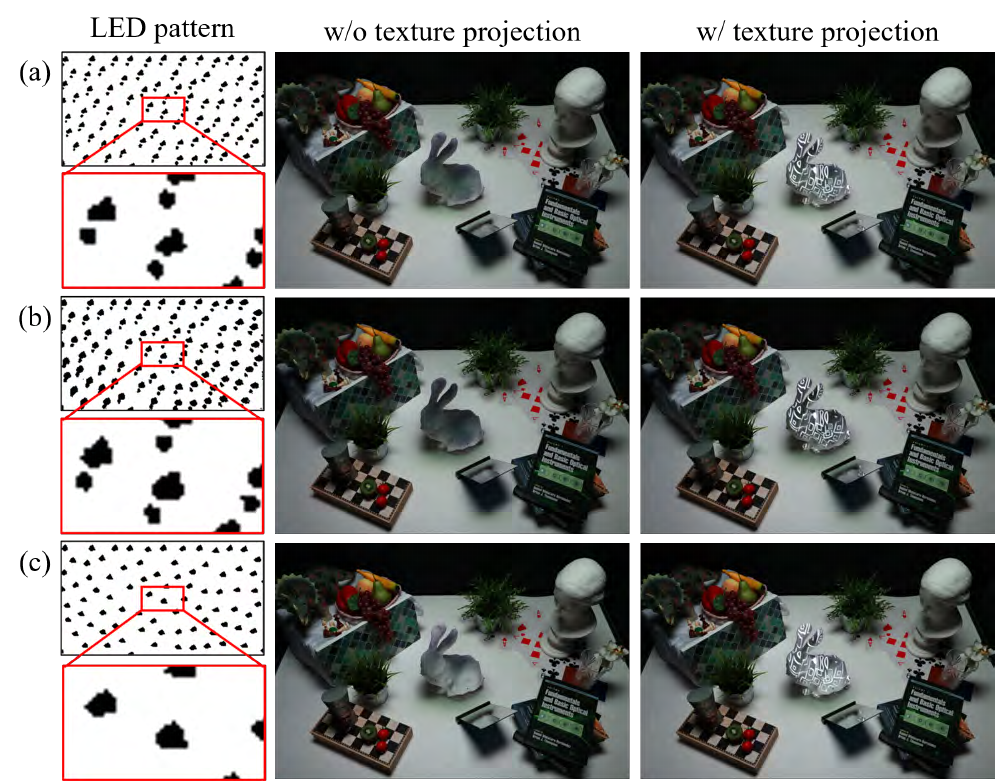}
    \caption{Comparison of three LED pattern computation methods under mirror reflections. (a) LTM-based and (b) optical ray tracing methods correctly suppress target illumination via mirror reflections, as indicated by the LED patterns where the pixels corresponding to the mirror region are turned off. (c) In contrast, the geometry-based method does not handle mirror reflections, leading to significant contrast degradation in the bunny’s foot region.}
    \label{fig:pm_results_With_Mirror}
\end{figure}

We also compared qualitative properties across the three computation methods.
We introduced a mirror into the scene and computed LED patterns with each method.
\autoref{fig:pm_results_With_Mirror} shows the resulting LED patterns, the scene under the target-excluding lighting, and their PM results.
The geometry-based method fails to account for mirror reflections and suffers a marked contrast drop in the bunny's foot region, whereas the LTM-based and optical ray tracing methods avoid illuminating the target via reflected paths and maintain contrast.
The LED patterns corroborate this: only the latter two methods intentionally darken small regions aimed at the mirror.

\begin{figure*}[t]
    \centering
    \includegraphics[width=\linewidth]{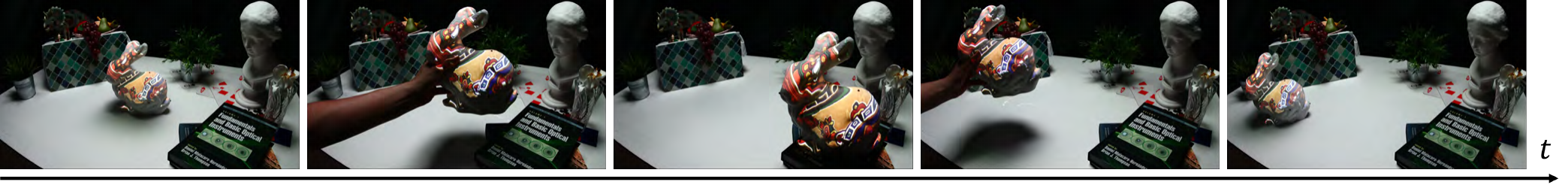}
    \caption{Dynamic PM on a moving target. Our system uses the geometry-based method to update the target-excluding lighting and texture projection in real time, maintaining high contrast throughout the motion.}
    \label{fig:dpm_results}
\end{figure*}

Finally, we measured computation times.
The LTM-based method, including acquisition and iterative optimization (gradient descent), required $\sim$12\,hours.
The optical ray tracing method required $\sim$7\,s from projecting white on the target to obtaining the LED pattern.
The geometry-based method ran in $\sim$16.6\,ms, corresponding to 60\,fps, indicating feasibility for dynamic PM.
Indeed, as shown in \autoref{fig:dpm_results} and the supplementary video, the system tracks target motion and maintains both target-excluding lighting and texture projection.

\subsection{Basic Characteristics}
\label{sec:EXPERIMENT_BASIC_CHARACTERISTICS}

We evaluate the five characteristics discussed in Section \ref{sec:RELATED_WORK}.
First, a sufficiently large effective light source area produces soft shadows under occlusion, leading to natural illumination.
We turned on all LEDs in our system and observed the shadow of a plaster bust on the floor (\autoref{fig:hardshadow_vs_softshadow}).
Compared with shadows produced by the projector of our system and by a typical LED luminaire (NEEWER NL660 Bi-Color LED Panel Light), the proposed system's shadow closely resembles that of typical lighting.
It can be observed that the shadows produced by the proposed system resemble those generated by typical lighting. Specifically, both result in soft shadows with blurred edges.
In contrast, projector-based illumination produces sharp-edged hard shadows.
From these results, we confirm that the effective light source area of the proposed system is sufficiently broad---comparable to that of typical lighting---and thus capable of reproducing natural illumination.

We evaluated luminous distribution using the beam angle.
In lighting, the beam angle is defined as the angular width at which intensity drops to 50 \% of the peak.
For our system, the beam angle was approximately 90 degrees.
Applying the same measurement to the typical LED luminaire yielded a beam angle of approximately 100 degrees.
These results indicate that our system offers a wide luminous distribution comparable to that of typical environmental lighting.

\begin{wrapfigure}{r}{0.25\textwidth}
    \centering
    \vspace{-15pt}
    \hspace{-20pt}
    \includegraphics[width=0.25\textwidth]{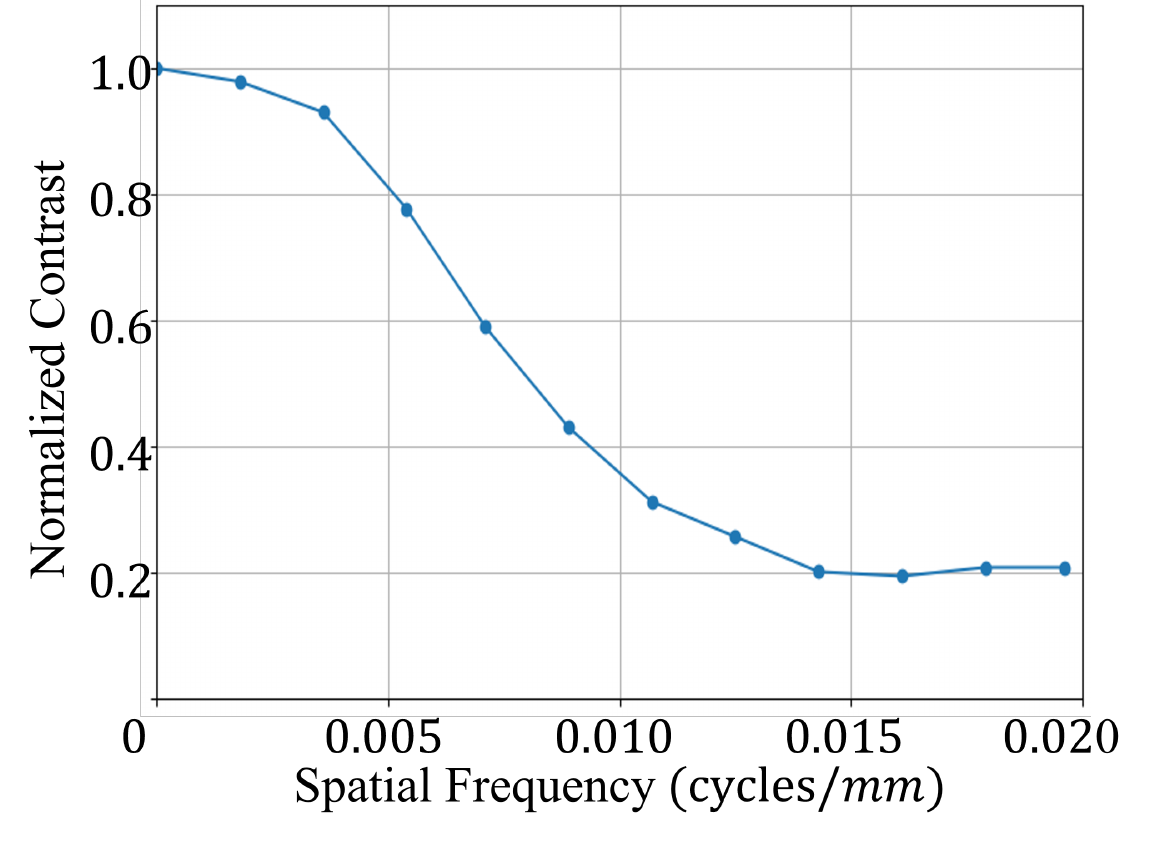}
    \hspace{-15pt}
    \vspace{-8pt}
    \caption*{\footnotesize{MTF of target-excluding lighting in the proposed system.}}
    \vspace{-8pt}
    \label{fig:mtf_result}
\end{wrapfigure}

Regarding spatial controllability, we evaluated it using the modulation transfer function (MTF), which represents the contrast of displayed images as a function of spatial frequency.
Specifically, we used the LTM-based method to generate an LED luminance pattern that projects a sinusoidal pattern onto the floor and calculated the MTF from the captured result (see the inset).
The results confirm that a contrast of 0.8 is maintained at a spatial frequency of 0.005 cycles/mm, which corresponds to a wavelength of 200 mm.
This suggests that for projection targets larger than 100 mm in width, it is possible to avoid illumination while suppressing the PM content's contrast degradation to within 20 \% of that in a dark room.
Furthermore, if a 50 \% contrast degradation relative to the dark-room condition is acceptable, the minimum excludable area width for this system is calculated to be about 60 mm.


For luminous efficiency, we compared illuminance measured at the floor center when all LED pixels were turned on under three conditions: without a lens array, with the periodic lens array, and with the proposed aperiodic lens array.
The measured illuminance values were 260\,lux (no lens), 180\,lux (periodic), and 130\,lux (aperiodic).
Thus, relative to the lensless maximum-efficiency case, the aperiodic array reduces efficiency by 50 \%, and relative to the periodic array by 72 \%.
Since the luminous efficiency of a typical DLP projector is reported to be 44 \%~\cite{SUN2011301}, the efficiency of the proposed system can be considered comparable to, or slightly better than, that of multi-projection-based systems~\cite{jones2014roomalive, fender2017meetalive, takeuchi2023projection}.
We confirmed that the primary factor reducing the efficiency of the proposed system is the lens aperture.
Previous systems that combined projectors with a lens array~\cite{yasui2024projection, cossairt2008light,yasui2021dynamic,yasui2021wide,zhou2015light,matusik20043d} or an LCD panel with a lens array~\cite{takeuchi2016anylight1,takeuchi2016anylight2,takeuchi2018integral, takeuchi2021theory} suffer from compounded losses: light loss in the projector or LCD panel, in addition to that caused by the lens array.
Consequently, their luminous efficiency is expected to degrade significantly compared with the proposed system.

For form factor, the combined LED display and lens array measured 800\,mm (W) $\times$ 420\,mm (D) $\times$ 200\,mm (H).
A height of 200\,mm is substantially thinner than projector-based systems, demonstrating an advantageous form factor.

\section{Discussion}
\label{sec:DISCUSSION}

In this study, we realized a novel target-excluding lighting by combining an LED display with an aperiodic lens array, achieving a thin form factor, inherently wide illumination coverage, and a large effective light source area that generates soft shadows.
Our experiments confirmed that the proposed lens placement optimization significantly suppressed unintended dark spots caused by crosstalk during target-excluding lighting.
With respect to spatial resolution, when the projection target exceeds a 100 mm cube, the contrast degradation remains limited to approximately 20 \%.
Furthermore, we verified that the spatial resolution of the proposed system was sufficient to selectively avoid projection targets larger than a 100 mm cube (approximately the size of a human fist), with the contrast degradation limited to approximately 20 \%.
We also confirmed that the proposed system improved the RMS contrast of projection mapping in bright-room environments from 0.148 to about 0.245 through target-excluding lighting.
Among the three LED luminance pattern computation methods proposed in this study, the two designed for faster calibration and processing showed almost no loss of contrast compared to the LTM-based method, which guarantees theoretical optimality, and the visual difference was negligible.
We discuss this point further in \autoref{subsec:limitation}.
To the best of our knowledge, this is the first demonstration of contrast enhancement in bright-room PM achieved even for dynamic PM onto moving objects.

\begin{figure}[t]
        \centering
        \includegraphics[width=\linewidth]{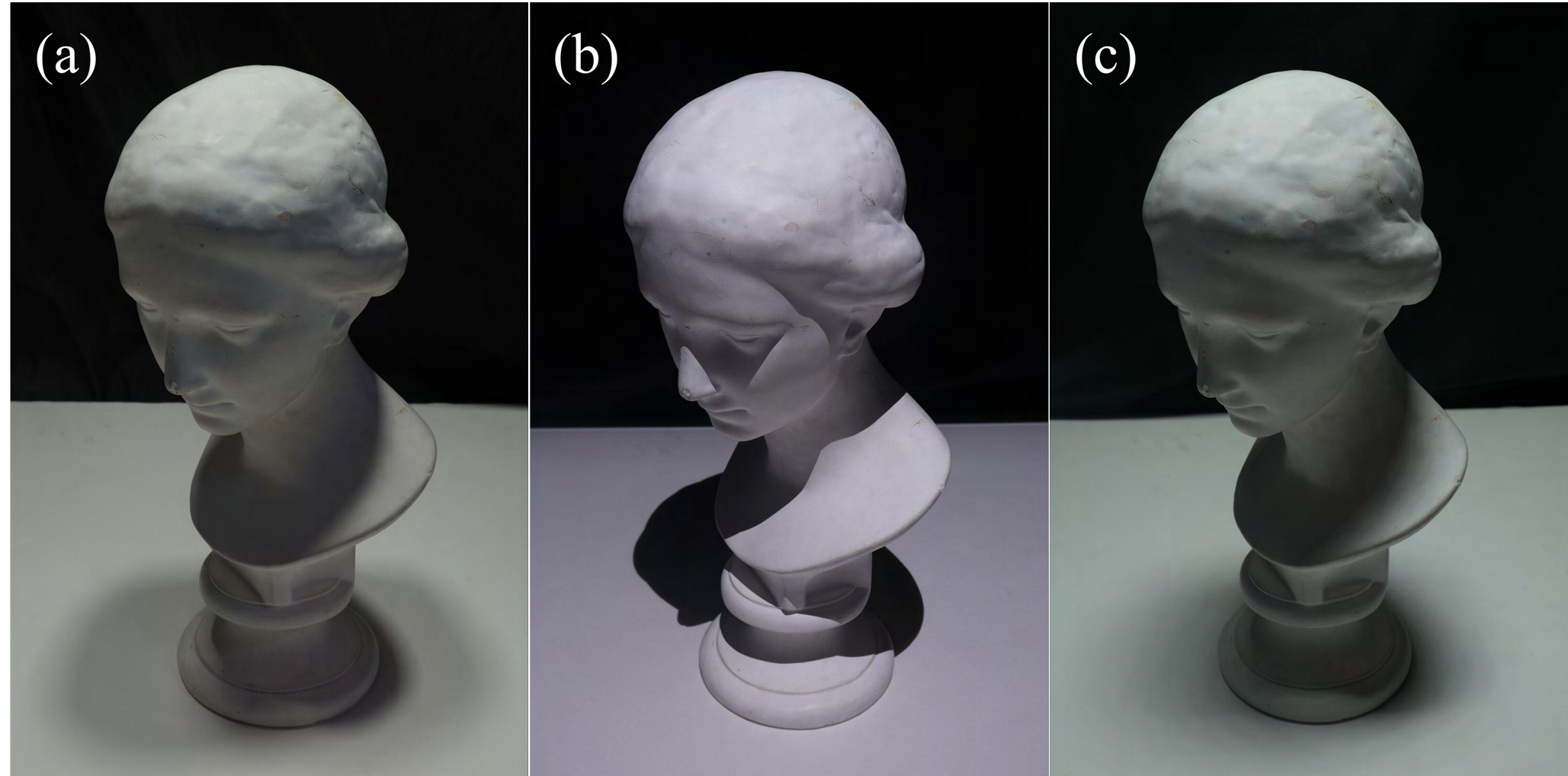}
        \caption{Comparison of shadows cast by (a) typical lighting, (b) a projector, and (c) the proposed system.}
        \label{fig:hardshadow_vs_softshadow}
\end{figure}

A previous work~\cite{takeuchi2024projection} revealed that enabling high-contrast projection mapping in bright-room environments shifts the perception of the projection target from an aperture color mode---where the target appears self-luminous---to a surface color mode, in which the target appears as though its reflectance properties have been altered.
Accordingly, our proposed technique similarly suggests the possibility of perceptual transformation into the surface color mode, that is, perception as a change in reflectance.
Achieving natural perceptual switching into the surface color mode has long been regarded as one of the central goals of PM research~\cite{10.1007/978-3-7091-6242-2_9}.
However, in the previous work~\cite{takeuchi2023projection,takeuchi2024projection}, target-excluding lighting was realized using multiple projectors, and shadows cast by environmental objects occluding the projector light were hard shadows, failing to achieve natural illumination.
By contrast, as illustrated in \autoref{fig:hardshadow_vs_softshadow}, our proposed technique produces soft shadows, thereby providing a more natural lighting environment.
As shown by prior vision science research~\cite{Mizushima:22}, the diffusivity of illumination is a critical factor in the perception of surface properties of illuminated objects.
With the strongly directional lighting of the previous work~\cite{takeuchi2023projection,takeuchi2024projection}, there is a risk that the surface properties of objects other than the projection target would not be perceived correctly.
Our proposed technique addresses this issue as well: it controls the appearance of the projection target in the surface color mode, as if its reflectance had changed, while maintaining the natural appearance of surrounding objects.
To the best of our knowledge, this research is the first work to achieve this.

\subsection{Limitation}
\label{subsec:limitation}

The proposed approach has several technical limitations.
First, target-excluding lighting illuminates the environment other than the projection target, while interreflections from these regions may brighten the target itself, particularly in areas close to environmental surfaces.
Even when optimizing LED luminance patterns using the LTM-based method, it is physically impossible to project negative light; thus, darkening the projection target region beyond a certain level is not feasible, and the difference in contrast relative to other methods was not substantial.
This limitation can also arise in previous target-excluding lighting techniques~\cite{takeuchi2024projection}, and it is difficult to resolve solely through lighting-side improvements.
Adjusting the brightness of projection content through radiometric compensation~\cite{grundhofer2015robust} or modulating the reflectance of the projection target~\cite{7383338} may mitigate this issue, which we consider an important avenue for future work.

Although adopting an aperiodic arrangement of lenses successfully prevented periodic dark regions caused by crosstalk during target-excluding lighting, compared with the closest periodic packing, the number of lenses that could be placed decreased.
As shown in Section \ref{sec:EXPERIMENT_BASIC_CHARACTERISTICS}, while 146 lenses could be used under closest packing, only 111 lenses were available with our optimized arrangement, resulting in an about 30\% reduction in light throughput.
To address this, we believe that allowing lens aperture, focal length, and even shape to vary during optimization could enable the design of lens arrays that maintain a large effective light source area while suppressing dark spots caused by crosstalk, making this an intriguing direction for future work.
Furthermore, employing gradient-based optimization, such as differentiable rendering, using the arrangement obtained by our greedy method as an initial value could potentially refine the lens placement to further increase density.

We also discuss the robustness of our method regarding target properties.
Regarding shape, while the LTM-based and Optical Ray Tracing methods handle complex geometries robustly on a per-pixel basis, the Geometry-based method approximates the target using a convex hull.
Consequently, for complex non-convex shapes, the excluded area may be slightly larger than necessary.
Regarding position, exclusion accuracy slightly decreases in the peripheral areas of the system.
This is because the projected LED images are magnified due to steeper incident angles to the lenses and longer projection distances.
Finally, regarding size, when the projection target becomes larger or when multiple targets are present, more LED pixels need to be turned off.
As a result, their crosstalk images reduce the illumination intensity of the surrounding environment.
Restricting the angular distribution of each LED to prevent crosstalk images would eliminate this problem, but it would also reduce the overall luminous distribution and effective light source area, thereby diminishing the lighting performance of the system.
Since this limitation is difficult to resolve with our system alone, it is necessary to consider system configurations that combine multiple techniques, such as multi-projector-based lighting~\cite{takeuchi2024projection}, to compensate for each method's weaknesses.
Investigating design methodologies that integrate multiple principles to realize target-excluding lighting remains an important challenge for future research.

\section{Conclusion}
\label{sec:CONCLUSION}

We realized a novel target-excluding lighting method by combining an LED display with an aperiodic lens array. The optimized lens arrangement significantly suppressed crosstalk-induced dark spots, and three LED luminance computation methods were introduced to balance accuracy and speed. Among them, the geometry-based method achieved real-time performance at 60 fps without notable contrast loss compared to the LTM-based method, demonstrating applicability to dynamic PM. Experiments showed that our system offers a far more compact form factor than prior multi-projection systems, while maintaining 50 \% luminous efficiency and an effective light source area comparable to typical lighting. Even for targets smaller than a 100 mm cube, contrast degradation was limited to under 20 \%, confirming highly effective target-excluding lighting. In future work, we aim to build the ultimate PM environment by integrating shadowless mechanisms~\cite{10049693}, making projection results indistinguishable from real objects illuminated by environmental lighting.

\acknowledgments{
The authors wish to thank colleagues in the XR Group, Graduate School of Engineering Science, The University of Osaka, especially Kazuyuki Miura.
This work was supported by JST BOOST, Japan Grant Number JPMJBS2402, Grant-in-Aid for JSPS Fellows under Grant Number JP25KJ1699, JST ASPIRE Grant Number JPMJAP2404, and JSPS KAKENHI under Grant Numbers JP25K03155 and JP25K22820.}

\bibliographystyle{abbrv-doi-hyperref}

\bibliography{template}
\end{document}